# On a thermodynamic basis for inflationary cosmology

Akinbo Ojo

*Standard Science Centre*
*P.O.Box 3501, Surulere, Lagos, Nigeria*
*Email: taojo@hotmail.com*
*Tel: 234-1-4345025*

**Abstract**

Modifying the standard hot big bang model of cosmology with an inflationary event has been very successful in resolving most of the outstanding cosmological problems. The various inflationary mechanisms proposed depend on the production of expansion from exotic phenomena, false vacuum, scalar fields or other exotic particle behavior within an environment of astronomically high energy, making such proposals relatively inaccessible for verification by experimental tests. Though descriptive of how space has been expanding, the models do not give a complete and consistent mathematical or physical explanation that compels space appearance and propels its expansion. Here we describe another mechanism for achieving exponential inflation based substantially on already tested physics and equations, particularly the thermodynamic equation, $\partial S = \partial E/T$ and relate this to the creation event.

**Key words:** Thermodynamics; Cosmology; Inflation

PACS Classification: 98.80.Bp, 98.80.Ft, 98.80.Cq

# I. Introduction

The belief in a creation event received significant impetus following Hubble's discovery of an expanding universe, which suggested a beginning, and this was further entrenched by the subsequent observation of the cosmic microwave background radiation. With support from the gravitational theories of the large scale as espoused by Newton and Einstein, indications from thermodynamics of a low or zero entropy beginning and the physical possibility of the emergence of a quantum of energy from absolute nothing as enunciated in the Heisenberg uncertainty principle, physicists now feel somewhat confident that a reasonable cosmological model can be designed. Following a short introduction, we briefly mention some pertinent deficiencies in current cosmologies in section II. We then give a simplified account of a thermodynamically based model in section III and make concluding remarks in section IV.

The initial standard hot big bang model has been improved upon by merging inflationary scenarios [1,2] with the hot fireball of the big bang. The inflationary models not only therefore preserve most of the successes of the original standard model but also in addition provide solutions to many other outstanding cosmological problems. The necessity for inflationary cosmology is seemingly indisputable from theory and available observational evidence. However most inflationary mechanisms proposed depend on the production of an acceleration of the scale parameter from exotic phenomena, false vacuum, scalar fields or other exotic particle behavior within an environment of astronomically high energy. Unfortunately, such highly energetic environments are inaccessible and cannot be easily built in particle accelerators for experimentation and so it is difficult to verify the correctness of these proposed mechanisms of expansion. While observationally, evidence may therefore objectively support inflation, we are left to speculate subjectively on the appropriate causative mechanism.

That thermodynamics must have some role to play in explaining the beginning and the subsequent evolution of the universe is a fact many are quick to recognize. In spite of this, the predominance of published work on the subject of the universe's appearance and subsequent expansion are based almost entirely on a gravitational framework with some added ideas from particle physics. To begin to grasp some of the essence of this paper, let us assume as a thought experimental (*gedanken experimenten*) situation that there were some other extra-terrestrial civilization who were yet to be acquainted with gravity and particle physics though well versed in thermodynamics (probably because while they already have their Boltzmann, their Einstein was yet to be born!), can they visualize like us a universal beginning through the thermodynamic lens? If an expansion of universal space is observed, e.g. from a reducing cosmic microwave background radiation temperature (to avoid the use of receding galactic clusters – which are gravitational masses!), can that civilization obtain some understanding of this? Can they describe an increase in universal space volume with time using the known thermodynamic equations? If there was indeed a beginning and as the second thermodynamic law seems to suggest, this was of possible zero entropy, what would this look like? Would this be a hot or cold state at zero kelvin as the third thermodynamic law somewhat suggests? If it were indeed a state at zero entropy and zero kelvin as the second and third laws seem to jointly suggest, what would be the effect of a quantum energy fluctuation occurring spontaneously in such a state if the energy quantum could exist for an astronomical amount of time according to the Heisenberg uncertainty relation?

Peering through an essentially thermodynamic lens therefore, the present paper tries as much as possible to describe and present what can be seen, if our known thermodynamic equations are largely correct and the Heisenberg uncertainty relation between spontaneous quantum energy fluctuations and time are useful without introducing much gravitational and particle physics based ideas and concepts like closed, open or flat space-time curvature, scalar fields, energy-momentum tensor, false vacuums, etc. This deliberate choice of view is not to claim that these other ideas are irrelevant to our complete understanding, neither is it to supplant the gravitational or particle physics view, but we try and describe as best as possible what can be seen using the thermodynamic window and hope that in further research there may or may not be correlation between the picture seen and what is described in current gravitational and particle physics inspired models.

Indeed, it may be asked, why bother looking at the universe through another lens when detailed cosmological models that incorporate gravitational and particle physics have been developed and these have been painstakingly tested against astronomical data with appreciable success. This is a valid question, but as it is known in spite of much expectation, the universe is yet to betray clear and unambiguous signs of obedience to the gravity that we can see by a decelerating expansion and cosmologists have had to resort to gravity and anti-gravity that we cannot see in the form of dark matter and dark energy to fill gaps and explain observed paradoxical behavior. For illustration, in the scenario of a beginning from a massive singularity, against common expectation the universe escapes the infinite gravity and furthermore after escaping instead of decelerating, the rate of expansion seems to necessitate a need for unseen matter to bring more gravity into the picture to explain a uniform or flat expansion or if as is currently speculated, there is an accelerating expansion in defiance of gravity, a dark energy is needed to rescue common expectations for gravitational behavior. Also some of the particle physics based ideas on inflation despite their success have been impossible to confirm experimentally and as a result several variations of the inflation scenario exist, many of these author dependent.

Although the universe can be regarded as a gravitational system, it is in many respects also a thermodynamic contraption. If it is "all there is" and nothing exists out of it, then it is a 'closed system', it also contains energy, entropy and a volume that serves as a compartment in which constituents can be arranged in various ways. Viewing cosmology with thermodynamic lenses may therefore really not be out of place. The hope of the paper is that the scenario described will be useful to complement our overall view of the universe in much the same way as radio-astronomy complements without supplanting optical astronomy, thereby giving us a clearer picture of the universe. Complementarity also implies that the thermodynamic scenario may answer some of the cosmological questions that the gravitational view finds difficult and the gravitational view too may in turn provide answers to areas that the thermodynamic view is hard pressed to explain. Such complementarity should help remove artefacts caused by the choice of 'lens' used and lead to a convergence of ideas on the remaining difficult areas in cosmology, some of which are briefly mentioned next.

## II. Some remaining conceptual deficiencies in the dynamics of current cosmological models.

The standard hot big bang model and the various proposed modifications by inflation are still plagued by some important deficiencies in their dynamical concepts. Apart from these deficiencies, there are also certain other issues that may not really be flaws but remain uncertainties. For example, it is uncertain whether there is any 'missing mass' to make the universe open, flat or closed and what consequences await the fate of the universe as a result. These flaws and the accompanying paradoxes could be useful in discerning artefacts in our view and provide pointers for improved dynamics in our cosmological models. Some of these are briefly discussed.

With gradual unanimity in definition, the universe is described as "all there is" and comprises not only the matter and energy forms but also all the empty space as well. That is, no matter, no energy and no space exists outside of the system we call the universe. Nothing out of it can influence it nor can anything be ejected out of it, making it a closed system. We also now know that it has not always existed. The first flaw now described arises from the manner of describing the emergence of space. This cosmological problem is also somewhat related to what is called the 'singularity problem' and part of it is that, where did space come from if it had not always existed? What equation or theory compels or mandates the emergence of space at time zero? What preceded this emergence, was it a state of nothing (i.e. no space, no matter, no energy, etc) or was it a massive, infinitely dense thing at zero radius?

The Friedmann-Robertson-Walker (FRW) models that are used to describe the picture in the standard big bang model and its particle physics inspired inflationary modifications are based essentially on the Einstein field equations. One equation that summarizes the dynamics well is:

$$R'' = -(4\pi G/3)\,(\rho + 3p)R \qquad\qquad (1)$$

where $R''$ is cosmic deceleration (and the prime $'$ denotes differentiation with respect to time), G is the gravitational constant, $\rho$ is the energy density, p represents pressure and R is the cosmic scale parameter (analogous to the radius of the universe in four dimensional space). The equation describes how the cosmic scale parameter R varies with time. As would be expected of gravitational behavior, as R increases universal expansion decelerates as Eq.(1) shows. When this equation is followed backwards in time, we reach a time zero when R was also zero representing the initial singularity of the big bang. When R = 0, R″ = 0 also, i.e. no acceleration and no deceleration at R = 0. One deficiency of the equation is therefore that it has no answer for us why R does not remain at zero and why space has to emerge at time zero. For the scenario of a universe starting from an infinitely dense concentration of matter and energy, a common proposal is that of an explosion occurring at time zero but Penrose [3] and others have noted that this cannot be an explosion of the ordinary kind in which material is ejected into some pre-existing space, rather the space itself is part of what has to be created at the beginning and it does not precede and exist as an arena for explosion to occur into. Similarly, the use of *high pressure* in the early universe as a mechanism 'driving' the expansion is inconsistent as noted by others, e.g. Rees, M.J. [4], since explosions are caused by *unbalanced* pressure and there was no 'empty' region or space outside the early universe for which pressure gradients might exist since the universe was and is 'everything'. Moreover in such a scenario, the envisaged high pressure should require collisions to occur between particles but it is assumed that the initial state was at zero radius without space between particles. We are therefore left with the deficiency of consistently explaining what brought about a change in R at time zero. Inflationary models based on gravitational and particle physics make use of various mechanisms to make the term (ρ + 3p) in bracket have a negative value in Eq.(1) above so that R″ becomes positive some later time after time zero, resulting in an acceleration of the scale parameter R with time instead of a deceleration. This gives a consistent explanation for exponential increase of already existing space. However, this still falls short of providing us with a mathematical and physically consistent explanation for the initial emergence of space right from time zero.

Newton's laws of motion are consistent for describing displacement from one place to another, either by force or by 'uniform motion in a straight line'. In the case of the universe however, there is no place outside of it for it to be displaced into either by force or by uniform motion, rather the universe itself is creating and increasing its own space within itself, out of nothing and cannot therefore be moved from 'place to place' in the common interpretation of 'motion'. That we encounter interesting paradoxes when we then try to describe the dynamics of the universe's motion in the usual way is therefore not surprising. In the same vein, viewing the expansion of the universe as work will not tally with the accepted definition because work implies a force moving something over a 'given' distance. However as mentioned, the expansion is not really in the common interpretation of a movement of material bodies through some displacement but a stretching of the existing distance between them by the emergence of more space, hence no real force is involved to cause movement and no work can be said to have been done. At a big crunch, we would similarly not expect movement in the common sense of it, like an apple falling from a tree to the ground under gravitational force. Rather the apple remains where it is and the ground remains where it is but the space between them shrinks gradually till it disappears completely! Therefore in that case as well no work may be done and no force may be needed for movement. These considerations imply that the observed reduction of temperature of the cosmic background radiation cannot therefore be consistently described as being a result of heat transfer nor of work done by the universe on its surroundings by its expansion but rather as a redistribution of the initial quantum of energy among universal space that is being further created.

For the case of the universe emerging from an infinitely dense massive singularity at time zero, the dynamical challenges include consistently explaining how matter would acquire a velocity higher than the velocity of light in order to move away from the singularity's center of mass, an impossibility according to Einstein's relativity. Yet the universe found a way to come into existence. For a universe starting from "nothing", clearly there will also be problems explaining the dynamics of the emergence in a way that would be compatible with Newton's laws of motion. Firstly, if momentum is zero from the beginning, from where do we conjure the initial momentum from, if the universe is 'all there is'? Secondly, there is the interesting difficulty that must arise in the context of Newton's third law of motion. The significance of the third law for a discussion of a cosmological beginning is that, even if there were an agency external to the universe that can provide a motive force to kick-start and initiate the universe's motion, the *action* from this agency is completely ineffectual because of the absence of a *reaction* to it, since the universe does not yet exist to provide this. From this basis as one instance, we may probably safely conclude that at least in the case of a universe starting from "nothing", the initial appearance of space was not due to the action of force. Following then from this, if force cannot initiate the universe's appearance from "nothing", and if it indeed emerged from "nothing", then a phenomenon other than force must describe the dynamics of its emergence and possibly its continuing expansion.

Although we intend to describe what can be seen from a thermodynamic viewpoint, a discussion of the dynamics of emergence and expansion will not be complete without briefly mentioning the cosmological parameter, omega ($\Omega$) and the significance of what its value at time zero portends for our cosmological modeling. $\Omega$ can be defined in a variety of fairly equivalent ways. For example, it can be expressed as: the ratio of the density of the universe to the amount of density required to close the universe and stop its expansion; a measure of space time curvature present in the universe – closed, flat or open; the ratio of gravitational energy to that of expansion energy; the ratio of the decelerating restraint to the accelerating impetus in universal motion, etc. Being an essentially thermodynamic perspective, the paper would favor the last of the equivalent ways of describing $\Omega$. When $\Omega < 1$, the universe is accelerating, when $\Omega > 1$, it is decelerating and when $\Omega = 1$, the universe is in uniform motion, if moving or at rest. Whatever is the chosen way of expressing $\Omega$, its current value appears not far from unity. This implies that $\Omega$ must have been incredibly close to unity in early eras for the current value to be within the vicinity of one after over ten billion years of expansion [4]. This however somewhat conflicts with the standard hot big bang model because in such a model of the universe starting from "something", the density of the universe at the beginning must have been so astronomical with $\Omega \gg 1$ and not the $\Omega \sim 1$ that our observations seem to indicate. This is part of the essence of the 'flatness problem' in cosmology.

Without an 'inflation scenario', the standard big bang model would face great difficulty explaining the observed value of $\Omega$ being close to unity in the early era if the universe started from a massive singularity and this problem was part of the motivation for introducing an inflationary scenario into the standard hot big bang model. Current inflationary proposals introduce a mechanism that causes an expansion some time soon after time zero bringing the density of the universe to a value that succeeds in giving us the acceptable value of $\Omega \sim 1$ in the early era, thus resolving the flatness problem. The models do not however explain the reduction in the universe's density in the period from time zero to when the inflationary epoch begins at the Planckian or lesser densities.

If the universe can on the other hand indeed start from nothing at time zero, without an impetus to accelerate it and without pre-existing gravity from an infinitely dense mass to decelerate it, i.e. accelerating impetus = 0 and decelerating restraint = 0, then acceleration is equal to deceleration, giving $\Omega = 1$ exactly to infinite precision at time zero. We may therefore conclude that if at time zero, $\Omega$ was nearly but not exactly unity, it must have started from "something". On the other hand if $\Omega$ at time zero was exactly one, then the universe must have likely started from "nothing", since a massive singularity would be of infinite density, giving $\Omega \gg 1$ at time zero before the inflationary epoch. It may be noted from Eq.(1) from which gravitational models are derived that when R = 0, there is zero deceleration and zero acceleration implying support for a value of $\Omega = 1$ at time zero. To avoid imposing *ad hoc* initial conditions, fine-tuning and the resort to the anthropic principle at time zero, this paper chooses to describe a universe that emerges from and is increasing from "nothing" which is seen as the more natural of the two possibilities, without the need for an accelerating impetus and without the initial presence of a decelerating restraint from the gravity of an infinitely dense singularity. Introducing a basis for the universe's expansion in spite of $\Omega$ being exactly one at time zero is part of what the submitted paper strives to put forward using known thermodynamic equations.

One final discrepancy in the standard hot big bang model needs mention before going ahead to section III and it is that, as also pointed out by Penrose, "a hot beginning with a gaseous fireball in expanding 'thermal equilibrium' will be inconsistent with the second law of thermodynamics which prescribes that in its initial state, the entropy of our universe was at some sort of minimum, not a maximum! And 'thermal equilibrium' is a term that refers to a state of maximum entropy"[3]. To illustrate with a familiar analogy, imagine witnessing a spreading wild forest fire. It will be natural to look back in time and correctly imagine that it must have started as a small speck of fire that has now spread. However looking further back still, we see that that initial burst of fire was preceded by a calm, 'cold' but probably highly inflammable state at time zero. Describing the very beginning as a 'hot' or 'cold' state may therefore depend on just how far back you look!

## III. A thermodynamically based model.

The universe started small, either from an infinitely dense thing at astronomical temperature or from 'nothing' [5-7] and Quantum theory and quantum events dominate the small scale. Many cosmological models suggest that a quantum fluctuation of energy against the background of a pre-creation state that was highly unstable may have heralded the creation event [5-9]. Describing creation using the laws of physics requires the spontaneous and uncaused appearance of something. Heisenberg's uncertainty relation initially applied to the pair, position-momentum but later found useful for the pair, energy-time, represents a tool that can serve this purpose and it has been previously exploited by cosmologists. A notable pioneer in the usage of the uncertainty relation as a cosmological tool was Edward Tryon [5] who visualized the possibility of the universe erupting out of nothing as a quantum fluctuation. According to Heisenberg's uncertainty principle such spontaneous quantum fluctuation of energy in a closed system is natural and does not require a cause, neither does it constitute a contravention of the laws of energy conservation, provided the energy so created from nothing returns back to nil within a given time as given by equation (2), making the phenomenon reversible.

$$\Delta E \times \Delta t = \hbar \qquad\qquad (2)$$

where $\Delta E$ stands for energy, $\Delta t$ = the time interval the energy fluctuation exists and $\hbar$ is $h/2\pi$, where $h$ is Planck's constant. The uncertainty relation has no energy limit but we may suspect that infinitesimal fluctuations will be more probable to occur than giant ones. Neither does the relation have a time limit, a fluctuation may exist for micro-seconds or for billions of years. For illustration, a quantum of energy of $10^{-52}$ joules can exist for more than ten billion years.

Although the prejudice for a pre-existing physical cause, agent or force to give impetus and kick-start space appearance and the continued expansion is understandable, it seems not much appreciated that if as is commonly conjectured, the universe is "all there is" and absolutely nothing was existing at the beginning, the very best we can hope to get in the circumstance will be a natural equation to describe the spontaneous physical appearance of space and its continuing increase thereafter. One advantage of equations predicting and describing the creation phenomenon is that it avoids the paradox of a "first cause or physical agent" but rather makes creation a matter of course. Furthermore, if the equations are devoid of conserved quantities and reversible then no relic is eventually left of the phenomenon, which may be of philosophical or aesthetic value. The mechanism now described appears devoid of the previously stated conceptual flaws, as the spontaneous physical appearance of space and its subsequent expansion is mathematically demanded without the requirement for a dynamical force. We now describe a mechanism for exponential inflation based on already tested physics and equations and as a result, it will not depend on the availability of giant particle accelerators for verification. It will depend only on the proviso that our current theories of gravity, quantum mechanics and thermodynamics are largely correct.

The following known thermodynamic equations, particularly $\partial S = \partial E/T$ (Equation 3 below), will be of prime importance in our description and have crucial roles to play cosmologically:

$$\partial S = \partial E/T \qquad\qquad (3)$$

$$S = k \log_e W \qquad (4)$$

From Equation (4),

$$e^S = W \qquad (5)$$

$$W = V^N \qquad (6)$$

The third law,

$$S \to 0, \text{ when } T \to 0 \qquad (7)$$

where, S = entropy, E = energy, T = absolute temperature, $k$ = Boltzmann's constant (here taken as equal to one for convenience), $\log_e$ = natural logarithm, $W$ = number of possible states that the system can assume, V = volume or number of compartments that can be occupied by the system and N = number of constituents of the system under reference. Lastly but not the least, we need to take cognizance of the second law before endeavoring to attempt a description of what can be visualized about our beginning through the thermodynamic lens. The law asserts that if a spontaneous change occurs within a closed system, the entropy in that system increases in time till an equilibrium state is attained, with entropy, S reaching a finite, maximal value. The entropy cannot increase beyond this finite value without a change in the parameters of the system. However it suffices to say that if an increase beyond this finite value is mathematically compelling and physically demanded, the system itself being a closed system must remove any barriers to higher entropy values and bring into effect any changes in its parameters to permit such physical or mathematically demanded increases in the finite maximal limit for entropy.

In the singularity theorems of General relativity (GR) [10,11], the initial singularity is described as that state from which space-time and matter emerge. In a sense therefore GR sees the beginning as a kind of absence of space-time and matter. In quantum theory, the fluctuations of energy in a closed system are not a conversion of one form of energy to another but the fresh appearance of a quantum of energy *de novo* according to Heisenberg's principle. Thermodynamically, the second law sees the beginning as one of likely zero entropy and the third law suggests that when entropy is about zero, the temperature of the state will likely be at zero kelvin, Eq.(7). Essentially therefore, these three disciplines of physics unanimously suggest one way or the other that the beginning was characterized by an absence of matter, space, energy, entropy and time. Philosophically it may appear more consistent to describe creation as the spontaneous emergence of matter, energy, space, time and entropy where these were hitherto absent rather than the conversion of an infinitely dense thing at infinite temperature to a less dense form.

The cosmological significance of the thermodynamic equation, Eq.(3), $\partial S = \partial E / T$ as a "creation equation" appears not to have received much consideration. This simple equation is unique in that it somehow relates the quantum and classical disciplines of physics believed to have a role to play in cosmology. These are thermodynamics ($\partial S$), quantum theory ($\partial E$) and gravitation (T, as a substitute for the proper-time of general relativity, because T$\rightarrow$ 0 kelvin, when proper-time $\rightarrow \infty$ as proposed in [12]. In space-time singularities, proper-time tends to infinity according to the singularity theorems of relativity). The equation $\partial S = \partial E/T$ is a reversible equation, i.e. when energy is completely removed from the system the entropy, no matter how astronomical it has become, reduces to zero. Significantly also the equation indirectly incorporates the enigmatic but cosmologically relevant concept of time through the Heisenberg principle that an initial quantum fluctuation, $\partial E$, will exist for a finite period of time.

Explaining the transition from an innocuous initial quantum occurrence to the classical picture of the universe we observe is at the center of cosmology. Eq.(3) having both quantum and classical concepts may therefore be uniquely useful in mathematically and physically describing this transition from a quantum fluctuation to an astronomical increase in universal space volume and entropy. The equation shows that if temperature were increased astronomically towards infinity, the increase in entropy caused by the addition of a quantum of energy will reduce to zero and such a state will be highly stable to the quantum fluctuations that may have given birth to the universe. Conversely, a state at absolute zero will be a highly unstable and volatile state. The appearance of the unimaginably tiniest quantum of energy results in monumental disorder when absolute temperature is zero kelvin. Although no time or dynamics is immediately obvious in Eq.(3), taking into cognizance the scenario at the beginning with T = 0 and the energy-time uncertainty relation that $\partial E$ must exist for a given time, time is incorporated and some dynamics will be inevitable because for as long as the time for a spontaneous energy fluctuation, $\partial E$ to exist has not elapsed according to the uncertainty principle, the entropy, S in a closed system must continue increasing towards some higher value in order to attain infinity. The equation vividly indicates the possible kind of scenario that would exist in a closed system such that a quantum fluctuation in energy can give rise to an astronomical sized increase in a classical parameter like entropy where this was previously absent. With the foregoing, we can now attempt a thermodynamically based model chronology of the creation event with appearance of space, matter, energy and entropy from a state where space, matter, energy, and entropy were non-existent.

In a singularity state at zero kelvin, a quantum fluctuation of energy results in an astronomical increase in entropy, see Eq.(3). The time duration for existence of this fluctuation is presumed governed by Heisenberg's principle. For as long as such a quantum of energy exists, Eq.(3) shows that the entropy of the state, S must continually increase towards infinity, giving us a possible glimpse as to the origin of the second law of thermodynamics. We have noted the cosmological inconsistency in attributing spontaneous appearance of space to the forceful agency of temperature or pressure. Now, Eq.(3) tells us that simultaneous with a spontaneous fluctuation in energy is a mathematically and physically demanded rise in entropy. Certainly, entropy cannot be manifest where there is no space, since $W$, the different possible arrangements the system can assume needs space for expression ($W = V^N$, Eq.(6)). With a demanded increase in entropy is therefore an exponential change in $W$ and space vacuum must necessarily make a physical appearance, even if none previously existed. Even if the quantum energy fluctuation was homogenous from the beginning, and this is possible since there was no space prior to the fluctuation in which heterogeneity may manifest, we see that it is not enough for entropy to have a maximal value at thermal equilibrium. This would be finite. In the scenario present at the beginning with absolute zero temperature, even if a finite value thermal equilibrium is present *ab initio* or is attained, entropy increase must strive beyond this finite limit in obedience to Eq.(3). One way a closed system that has already attained thermal equilibrium can further achieve an increase in its own entropy is by increasing the compartmental volume available to the system. The second law of thermodynamics seems to inform us that this first moment of creation is continuing, with S and V still increasing, observable as the expansion of the universe and the continuing increase of entropy with time. The dynamics of the scenario become clear if we ask, given the conditions at the beginning and what the equations imply, is it possible for entropy, S to be increasing while mathematically the compartment volume, V remains static? V can remain static, as S increases till equilibrium in a closed system but where S has to surpass the initial finite limit of thermal equilibrium, higher values of entropy and equilibrium can only occur by effecting changes in the compartment volume of the system.

Considering Eqs.(4),(5) and (6) and differentiating with respect to time t, we may write,

$$\partial S / \partial t \propto \partial (\log_e V) / \partial t \qquad\qquad (8)$$

or alternatively stated,

$$e^{\partial S / \partial t} \propto \partial V / \partial t = \partial (4\pi R^3 / 3) / \partial t \qquad\qquad (9)$$

We know from the second law that universal entropy is increasing with time but we do not know the rate of this increase. Eq.(8) or (9) seem to imply that if we know the rate at which entropy is increasing according to the second law of thermodynamics, we can deduce the rate of expansion as well, and vice-versa with the rate of change of entropy being proportional to the rate of change in the logarithm of the compartment volume. If S increases by arithmetic progression, the compartment volume, V increases by geometric progression and if V is increasing at constant radial velocity by arithmetic progression, the rate of increase of S will be declining with time. While Eq.(3), $\partial S = \partial E / T$, serves to guide us about the initial conditions present at time zero, Eqs.(8) and (9) give us the dynamical framework with which we may observe and monitor the expansion. In view of the possibilities for distortion from forces, like gravity acting against the unencumbered spreading of galactic clusters, astronomical observations of the rate of reduction of cosmic microwave background density should be able to serve as a fairly good monitor and give us quantitative estimates of the rate by which V is increasing, whether geometrically, at a constant rate or at a declining rate.

As the equations show, at time zero just before the quantum fluctuation occurs, E, S and V are zero, with the exponential increase of space starting at time zero. With $S \neq 0$, $W > 1$ and $V \neq 0$, then N too can no longer be zero ($N \neq 0$, see Eq.(6) ) and therefore constituents must appear in the system simultaneously, whatever they may be made up of. As is however common with nature, the appearance of space vacuum to serve the system for its compartmental properties may for economy be created in quanta to equally serve as the constituents, N of the system of the early universe. Already current quantum gravity proposals strongly speculate that like matter and energy were found to be expressed in discrete units in the kinetic and quantum theories respectively, space too appears to be granular at some infinitesimal scale, whether described as 'quantum foam', 'beaded strings', 'ether', etc. It remains a matter of speculation and future discourse whether these same constituents, N can be the fundamental structural building blocks for matter in highly energetic environments.

Although this granular space nature seems to conflict with the picture of continuous space in classical gravity, this may be more of another kind of duality (cf. wave-particle duality). If space is considered as the embodiment of what exists, then between space units "no space" exists, which is paradoxically still the same thing as saying space is 'continuous'. Geometrically, the quantization of space confers the Euclidean point with the finite dimensionality that removes the paradox in the idealized 'point' of zero dimension realistically conferring dimension to 'lines', 'surfaces' and 'bodies'.

We may therefore conceptualize 'universal expansion' as a thermodynamic increase of the compartment volume of the system. From Eq.(3), which is a reversible equation and the Heisenberg principle we also see that what could possibly determine how long the universe exists may be the time interval that has been imprinted on it by the uncertainty relation and from the thermodynamic perspective 'missing mass' may possibly not be needed to close it.

That initial quantum energy fluctuation can be assumed to be now occupying a larger volume and we may conjecture this to be the cosmic background radiation. The fact that the background radiation is very uniform and presumably at thermal equilibrium from the beginning till the present time and yet this does not seem to hinder the further increase of universal entropy seems a possible support of Eq.(3) and the scenario that may have existed at time zero, as earlier discussed. Other energy forms that undoubtedly exist in the universe may be speculated to have been derived from the granular space vacuum that has been created and may be classified as positive energy (matter and fields) and negative energy (gravity), with a possible net value of zero as often speculated [5-7], thus not conflicting with laws of energy conservation.

The continuous increase of space cannot be confined only to that between galactic clusters but must be universal since space is everywhere, including within the atom. There will thus be a pervasive tendency for organized structure to disintegrate and spread as a result (a common manifestation of the second law of thermodynamics). Where bonding is not strong, actual spreading will occur (viz. Hubble expansion). Where bonding is stronger, it is our speculation that orbits will be created with a delicate balance between the binding force and the tendency to spread. Although quantum physics gives us a reliable solution for atomic stability, using the stability of the simplest hydrogen atom for theoretical speculation, an electron will travel the $5 \times 10^{-11}$ meter radius and collapse into the nucleus under electromagnetic force of attraction in $\sim 10^{-17}$ seconds, unless the space between the electron and the nucleus stretches at least by similar amount. If atomic stability can indeed be related to expansion of space, we can theoretically use this as a quantitative estimate of the rate of radial expansion of space, which must then be $>10^6 \text{ ms}^{-1}$.

Another implication of the relationships in Eqs.(8) and (9) is that no matter how V is increasing, although total universal entropy too will be increasing, the entropy per unit volume will be reducing with time. That is, the number of different ways the constituents, N can be arranged within a certain volume of the system is decreasing with time. The only way this can be achieved is for some of the constituents of that volume to become thermodynamically unavailable to be arranged in a random way (cf. inertia in the gravitational view). Exhibition of such exclusion from rearrangement would seem manifest as foci of stability or structure in the system. and Eqs. (8) and (9) seem to suggest that this scenario would be increasing with time as the ratio of entropy to volume drops. According to Eqs.(8) and (9) the number of such constituents excluded from random distribution and spreading should be increasing as entropy per volume drops (cf. sub-atomic particles→ atomic structure→molecules →gaseous

aggregations→stars→ galaxies→ galactic clusters, etc). This much may be discernible through the thermodynamic lens, speculations on the nature of such aggregations of the constituents, N and the interactions between them and whether they acquire features such as mass and charge to facilitate the thermodynamic requirement for further clumping and unavailability for random distribution may be more appropriate for gravitational theory and particle physics. However, an observation that some of the constituents of a system are confined and constrained from being randomly distributed while others remain free to be arranged in any of a variety of ways must in some yet to be fully understood way give rise to the idea of force. Further theoretical speculations and experimental research to fill some other cosmological gaps, particularly that of whether our physical 'constants' are time-varying or not, will be necessary to give a more comprehensive picture of how our universe began small, most probably from nothing and became so big.

## IV. Concluding remarks

The scenario painted strongly buttresses inflationary modifications to the standard hot big bang model based essentially on thermodynamics and few already widely accepted experimental, theoretical and mathematical assumptions. Like the particle physics inspired inflationary models, the success of the standard model is also preserved because there is an initial astronomically high concentration of energy sometimes in the past. Aspects of inflation theory supported include the suggestion of a cold, even if inflammable beginning preceding the hot phase and the strong speculation that all the matter, energy, entropy and space were created from a beginning where there was no energy, no matter, no entropy and no space, making the universe probably devoid of any conserved quantities. The successes of inflationary cosmology in resolving cosmological difficulties by an inflation event are hopefully largely retained because the thermodynamic equations confirm the inevitability of physical emergence and astronomical increase in space if a quantum fluctuation occurs in a singularity state at absolute zero, making "inflation" a natural, inevitable and expected phenomenon rather than a contrived one. From the thermodynamic perspective it is possible to speculate solutions to the 'horizon problem' and why there is remarkable homogeneity of the cosmic background radiation, since there was no space at the beginning for any non-uniformity in the quantum energy fluctuation to manifest. 'Flatness' at the beginning with $\Omega = 1$ exactly is natural because all may have started from nothing, where there was no pre-existing motive force for acceleration nor a pre-existing massive body that could provide a decelerating gravitational force. However it would appear that the current rate of expansion may not be theoretically predictable but will have to be determined experimentally. Eqs.(8) and (9) can however accommodate both a constant rate of expansion ($\Omega = 1$) and an accelerating expansion ($\Omega < 1$). Monopoles and any 'missing mass' cannot be seen with thermodynamic lenses, only by Grand Unified Theories.

As a further addition to current inflationary proposals, the thermodynamically based scheme here presented answers a few additional questions. The inflationary mechanism does not require the presence of a force from high temperature or pressure for space to emerge and expand which will appear an inconsistency and paradox if the universe were 'all there is', as had been earlier noted. Unlike most inflation models, where the exponential inflation starts some fractions of a second after time zero, the inflation here starts right from time zero as commanded by the equations. Again unlike some inflation models but like the standard big bang model, the initial quantum energy fluctuation is the same as the 'primordial fireball' of the big bang and not a fireball developing after inflation to cause the big bang as some particle physics inspired inflation models propose [1,2]. That entropy is now so large although this has not always been the case follows from the thermodynamic equations and the probable scenario at time zero. Also why increases in entropy and space expansion are both in the same forward direction of time and the possible origins for the second law of thermodynamics can be hopefully contemplated. Origin of structure is not easily visible with thermodynamic lenses but what may be the fundamental building block for substance has been speculated. Origin of structure may be more appropriate speculative areas for gravitational theory and particle physics and present modeling requires matter-antimatter asymmetry to somehow occur in early cosmic evolution to give rise to all the matter surviving in the universe now. A final speculation is a possible quantum mechanical mechanism that can determine the life span of the universe based on the uncertainty principle and in spite of $\Omega \leq 1$ or having any other value. We submit finally that our current theories and experimental observations, particularly thermodynamics provide a consistent basis for inflationary cosmology.

## Acknowledgement

I thank Prof. Animalu of the Nigerian Academy of Science and Dr.Gboyega Ojo of the University of Lagos for their encouragement. And not forgetting Hrvoje Nikolic for help rendered.